\documentclass[%
 aip,
 amsmath,amssymb,
 reprint,%
]{revtex4-1}

\usepackage{graphicx}
\usepackage{dcolumn}
\usepackage{bm}
\usepackage{soul}
\usepackage{xcolor}

\usepackage[utf8]{inputenc}
\usepackage[T1]{fontenc}
\usepackage{mathptmx}
\usepackage{etoolbox}

\newcommand*\tageq{\refstepcounter{equation}\tag{\theequation}}
\usepackage{amsmath}
\usepackage{amssymb}
\usepackage{upgreek}
\usepackage{subcaption}
\usepackage{soul}

\makeatletter
\def\@email#1#2{%
 \endgroup
 \patchcmd{\titleblock@produce}
  {\frontmatter@RRAPformat}
  {\frontmatter@RRAPformat{\produce@RRAP{*#1\href{mailto:#2}{#2}}}\frontmatter@RRAPformat}
  {}{}
}%
\makeatother
\begin{document}

\preprint{AIP/123-QED}

\title{Radiation-dominated injection of positrons generated by the nonlinear Breit-Wheeler process into a plasma channel}
\author{Dominika Maslarova}
 \affiliation{Institute of Plasma Physics of the Czech Academy of Sciences,
Za Slovankou 1782/3, 182 00 Prague, Czech Republic}
 \affiliation{Faculty of Nuclear Sciences and Physical Engineering,
Czech Technical University in Prague, Břehová 78/7, 115 19 Prague, Czech Republic}
 \email{maslarova@ipp.cas.cz}
\author{Bertrand Martinez}%
\affiliation{ 
GoLP/Instituto de Plasmas e Fusão Nuclear, Instituto Superior Técnico, Universidade de Lisboa, 1049-001 Lisbon, Portugal
}%
\author{Marija Vranic}%
\affiliation{ 
GoLP/Instituto de Plasmas e Fusão Nuclear, Instituto Superior Técnico, Universidade de Lisboa, 1049-001 Lisbon, Portugal
}%

\date{\today}

\begin{abstract}
Plasma acceleration is considered a prospective technology for building a compact multi-TeV electron-positron collider in the future. The challenge of this endeavor is greater for positrons than for the electrons because usually the self-generated fields from laser-plasma interaction are not well-suited for positron focusing and on-axis guiding. In addition, an external positron source is required, while electrons are naturally available in the plasma. Here, we study electron-positron pair generation by an orthogonal collision of a multi-PW laser pulse and a GeV electron beam by the nonlinear Breit-Wheeler process. We studied conditions favorable for positron deflection in the direction of the laser pulse propagation, which favors injection into the plasma for further acceleration.
We demonstrate using the OSIRIS particle-in-cell framework that the radiation reaction triggered by ultra-high laser intensity plays a crucial role in the positron injection. It provides a suppression of the initial transverse momentum gained by the positrons from the Breit-Wheeler process. For the parameters used in this work, the intensity of at least $2.2\times 10^{23}~\mathrm{W/cm^2}$ is needed in order to inject more than 1\% of positrons created. Above this threshold, the percentage of injected positrons rapidly increases with intensity.  
Moreover, subsequent direct laser acceleration of positrons in a plasma channel, using the same laser pulse that created them, can ensure a boost of the final positron energy by a factor of two. The positron focusing and guiding on the axis is provided by significant electron beam loading that changes the internal structure of the channel fields.
\end{abstract}

\maketitle

\section{Introduction}
Positron acceleration in plasmas has gained great interest in recent years due to its promising future applications, such as a~multi-TeV electron-positron collider \cite{blue2003plasma}. Tremendous progress in the development of plasma-based electron accelerators in terms of high-quality femtosecond few-GeV electron beams has been achieved \cite{brunetti2010low, plateau2012low, weingartner2012ultralow, couperus2017demonstration,li2017generation,leemans2006gev,wang2013quasi,shin2018quasi,gonsalves2019petawatt, lundh2011few,buck2011real,zhang2017femtosecond}. However, the acceleration of positrons still lags behind electrons by around a decade of research \cite{joshi2020perspectives}. The advance is slower mainly due to the following two reasons: 1) external positron source is required \cite{albert20212020} and 2) the accelerating structures generated by a~particle or laser driver in plasmas are usually defocusing for positrons \cite{joshi2020perspectives,schroeder2010physics}.

The production of a positron beam for the $e^{-}e^{+}$ collider can be established when a relativistic electron beam from a conventional accelerator impinges on a thick
solid target \cite{clendenin1989high}. However, there is also an option to generate positrons with cutting-edge laser facilities.
The interaction of a highly intense electromagnetic field with a matter under the right conditions results in electron-positron pair creation. In order to spontaneously generate pairs in a vacuum, the field of value $E_c=m_e^2c^3/(e \hbar)\sim 10^{18}~\mathrm{V/m}$ is required, where $m_e$ is the electron/positron mass, $c$ is the speed of light in vacuum, $e$ is the elementary charge and $\hbar$ is the reduced Planck constant. This value, also known as the Schwinger field, for a typical experimental laser wavelength $1~\mathrm{\upmu m}$, corresponds to intensities in the range of $10^{29}~\mathrm{W/cm^2}$. Such a magnitude is out of technological reach in near future. 

Nevertheless, the pair generation can be also induced by different approaches, e.g. by the collision of an intense laser field with a relativistic electron beam, which can experience the Schwinger field in its rest frame. The relevance of quantum effects on electron/positron dynamics is typically described by the dimensionless quantum parameter $\chi_e=|p_\mu F^{\mu \nu}|/(m_e c E_c)$, where $p_{\mu}$ is the electron/positron four-momentum and $F^{\mu \nu}$ is the electromagnetic tensor. When $\chi_e\gtrsim1$, the particle experiences the electric ﬁeld $E \gtrsim E_c$ in its rest frame, and the pair creation is nonnegligible.
Positron generation by the laser-electron beam collision was demonstrated in the proof-of-concept SLAC E144 experiment in 1997 \cite{burke1997positron}. This experiment used the intensity of $\sim10^{18}~\mathrm{ W/cm^2}$ and electron ener\-gy~of~46.6~GeV, reaching $\chi_e \sim 0.1$, which generated $\sim 100$ positrons. The increase in the laser power would naturally lead to higher positron numbers. 

The ongoing development of multi-PW laser facilities \cite{5501061,sung20174, zou2015design, hernandez2010vulcan, meyerhofer2014omega,kawanaka2016conceptual, peng2021overview} is expected to become a new milestone in the physics of laser-matter interaction, bringing completely novel and unexplored insight into quantum electrodynamics (QED) processes. This also opens possibilities for laboratory astrophysics to study QED events naturally occurring in space, for instance in the magnetospheres of pulsars \cite{michel1982theory}.
As a consequence, there has been an immense motivation to study and propose possible schemes for generating positrons with extreme laser fields.
A significant yield of positrons reached by a collision of a laser pulse with a relativistic electron beam has been demonstrated in various theoretical and simulation works \cite{blackburn2014quantum,lobet2017generation,vranic2018multi}. 

\begin{figure*}
\centering
\includegraphics[scale=1]{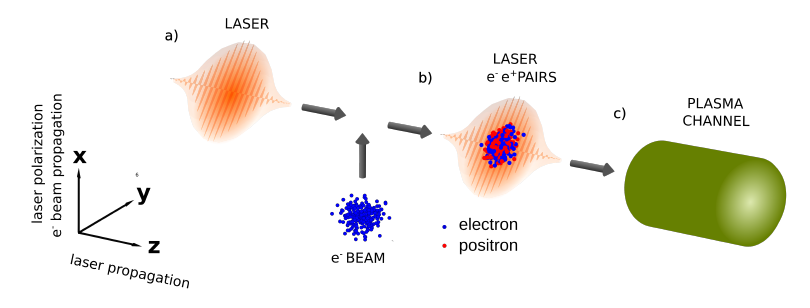}
\caption{Setup of the positron generation and acceleration: a) The laser pulse is moving in the $z$ direction and is polarized in the $x$ direction. It is colliding perpendicularly with the electron beam moving in the $x$ direction (blue dots). b) Consequently, electron-positron pairs are generated by the nonlinear Breit-Wheeler process, preceded by the nonlinear Compton scattering. A fraction of the pairs is deflected in the $z$-direction (blue and red dots in the central part of the laser pulse). c) Positrons that are deflected in the $z$ direction can be further directly accelerated in the plasma channel (green) by the laser pulse. }\label{fig:setup}
\end{figure*}
Moreover, the generation of positrons via the Bethe-Heitler process \cite{heitler1954quantum} by irradiating solid targets with a single laser pulse has been achieved experimentally and in numerical simulations \cite{chen2009relativistic, chen2010relativistic,  liang2015high, chen2015scaling}. The pairs can also be provided by an electron beam from a laser wakefield accelerator impinging on a solid target \cite{sarri2013table}.
As demonstrated by simulations, one can create pairs via the Breit-Wheeler process using $\sim10~\mathrm{PW}$ laser powers and solid targets \cite{ridgers2012dense,gu2018brilliant}. Furthermore, it has been proposed that two \cite{nerush2011laser, chang2015generation,zhu2016dense,jirka2017qed} or multiple \cite{vranic2017electron, gong2017high,efimenko2018extreme}
intense laser pulses can be effectively coupled for pair production when interacting with matter. The influence of the target density and length on pair production by the linear and nonlinear Breit–Wheeler and Bethe–Heitler processes has been also recently extensively explored, showing that the relative yield of each mechanism depends on these parameters\cite{he2022achieving}. In addition, the angular momentum of a positron beam can be controlled if its generation is established by specially tailored counter-propagating laser pulses, e.g. by two circularly-polarized pulses \cite{zhu2018generation} or a twisted laser pulse colliding with a Gaussian laser pulse \cite{zhao2022all}.

Regarding the second challenge, i.e. the focusing and acceleration of positrons in plasmas, most of the structures beneficial for electron acceleration, such as a nonlinear wakefield wave created by a laser or particle beam driver, are defocusing for positrons \cite{joshi2020perspectives,schroeder2010physics}. However, recent research has proven that the plasma is capable of providing both sufficient acceleration and focusing for positrons.
This has been experimentally demonstrated by the generation of plasma wakefield \cite{blue2003plasma, corde2015multi, gessner2016demonstration, doche2017acceleration, lindstrom2018measurement}, where a positron beam was used as a driver. 
Many theoretical and simulation works have recently proposed various schemes for positron acceleration as well. For instance, hollow electron beams \cite{jain2015positron} or hollow Laguerre-Gaussian "donut" laser beams \cite{vieira2014nonlinear} can focus positrons on the propagation axis. 
A combination of Gaussian and hollow laser beams can also enable positron injection if a slice of preformed plasmas contains positrons \cite{xu2020new}. Another possibility of providing a guiding structure is to tailor background plasma density to a finite-radius column \cite{diederichs2019positron} or a hollow plasma channel \cite{gessner2016demonstration, lindstrom2018measurement, silva2021stable}.
An alternative way of accelerating positrons in the nonlinear laser wakefield is tail-wave-assisted positron
acceleration, where the laser intensity is slightly above the threshold for the nonlinearity. In this case, some electrons can form an on-axis ﬁlament in the second period of the wakefield, creating both accelerating and focusing spot for positrons \cite{liu2023tail}.

Nonetheless, all these schemes still require a positron source based on conventional technology. The ability to generate $e^{-}e^{+}$ pairs by highly intense laser pulses and effectively accelerate positrons in plasmas can be, however, harnessed to build a one-stage plasma-based positron accelerator. Recently, an arrangement where the pair production is initiated inside a plasma wakefield wave, which provides the instantaneous acceleration, has been proposed\cite{liu2022trapping}. Moreover, it is possible to achieve linear Breit-Wheeler pair-creation in a structured overdense target \cite{he2021dominance, he2021single}, where the plasma magnetic ﬁeld can enable subsequent conﬁnement and direct laser acceleration of the pair-originated positrons \cite{he2021dominance}.

In this paper, we discuss the ability to use a single highly intense laser pulse to both generate and accelerate positrons. In particular, we analyze how many of the positrons created during an orthogonal collision of the laser pulse and an electron beam are deflected in the direction of the laser pulse propagation. Such positrons that propagate further forward with the laser pulse, within the area of its focus, can be promptly accelerated in plasmas. Therefore, the efficiency of the positron deflection and focusing on the axis reflects the potential of using this one-stage technique.
Such a setup has been already proposed in our former work \cite{martinez2023creation} (see Fig.~\ref{fig:setup}), where the features of the plasma acceleration were discussed in detail. Here, we mostly focus on the initial deflection part of the arrangement.

The generation and acceleration process can be initiated as follows. Firstly, an orthogonal collision of a relativistic electron beam with an intense laser pulse produces gamma photons through the nonlinear inverse Compton scattering \cite{vranic2018multi} (Fig.~\ref{fig:setup}a). The interaction of the intense laser light with the gamma photons afterward results in the generation of electron-positron pairs.
In contrast to the counter-propagating geometry, where the quantum parameter of an electron from the electron beam is $\chi_e\sim 2\gamma a_0\hbar \omega_0/(m_ec^2)$, the parameter will be twice as small in the perpendicular geometry:
\begin{equation}\label{chi}
\chi_e\sim \gamma a_0\hbar \omega_0/(m_ec^2). 
\end{equation}
\noindent Here, $\gamma$ is the relativistic Lorentz factor of the electron, $\omega_0$ is the laser frequency and
$a_0=0.855 \lambda_0[\mathrm{\upmu m}] \sqrt{I [10^{18}\mathrm{W/cm^2}}]$ is the normalized vector potential, where $I$ is the laser intensity and $\lambda_0$ is the laser wavelength. In principle, the orthogonal geometry leads to a lower value of the quantum parameter and, thus, a lower number of $e^- e^+$pairs. However, the setup profits from the fact that some of the particles can be caught in the phase with the laser field and get accelerated in the direction of the pulse propagation \cite{vranic2018multi} (Fig. \ref{fig:setup}b). This brings the possibility of boosting the final energy of the pairs. 
Moreover, if a plasma channel is placed directly behind the area of collision/pair generation (Fig. \ref{fig:setup}c), the positrons can be accelerated by direct laser acceleration~(DLA) mechanism \cite{pukhov1999particle} enabled by modified channel fields. In the typical DLA process, the combination of the transverse electric and magnetic fields of the laser pulse and channel focuses electrons on the axis \cite{hussein2021towards}. As a result, it generally acts against the positron guiding. However, an excess of negative charge can be progressively built up at the channel center, inducing a negative charge separation field, which can overcome the original self-generated field of the plasma channel, establishing on-axis positron focusing \cite{martinez2023creation}.

Previous work discussed the transition between steps b) and c) in Fig. \ref{fig:setup}\cite{vranic2018multi}. In particular, the conditions that are favorable for the positron deflection in the direction of the laser pulse propagation were analyzed. The results showed that particles with lower initial momentum (which is nonzero in the direction of the seed $e^-$ beam propagation), are deflected with much higher probability. This is caused by the fact that the maximum possible energy of created pairs is inversely proportional to their initial Lorentz factor $\mathcal{E}_{max}\sim\gamma_0^{-1}$ because $\gamma_0$ influences the dephasing between the particle and the wave. However, in the case of multi-PW laser pulses, radiation reaction (RR) can significantly affect particle dynamics. Radiation reaction means that the radiation emitted by a positron acts back on its motion.
Its damping effect can be approximated classically for $\chi_e \lesssim 0.1$, while for $\chi_e\gtrsim0.1$, QED stochastic effects become visible. For $\chi_e\gtrsim 1$, particles radiate a significant fraction of their energy.

The RR damping effects in high-intensity fields have been shown previously in simulations \cite{ji2014radiation,vranic2018extremely,gong2019radiation,yeh2021strong}. If electrons from an initially overdense target interact with an impinging strong-field laser pulse, the electrons are transversely trapped instead of being pushed outward. This radiative trapping is caused by the RR force, oppositely directed to the ponderomotive force \cite{ji2014radiation}. A similar effect has been observed during the formation of an underdense plasma channel\cite{vranic2018extremely}.
Moreover, RR can actually enhance the energy gain of electrons from a laser pulse in a strong magnetic field in a dense plasma. As transverse electron momentum is reduced through friction, the electrons can be accelerated more efficiently and gain more energy \cite{gong2019radiation}. In addition, if the wavefronts of a tightly focused laser pulse are superluminal, electrons with various initial energies reach roughly the same maximum energy due to the interplay between the superluminosity and RR\cite{yeh2021strong}. RR, therefore, needs to be taken into consideration in the interaction of particles with an intense laser pulse, such as in our setup presented in this paper. The impact of RR on the positron deflection and acceleration will be the main subject of this study.

This paper is structured as follows: in the second section, we theoretically analyze the motion of a newly created positron in a strong-field electromagnetic plane wave. We compare the derivation without radiation reaction and with classical radiation reaction approximated by the Landau-Lifshitz equation. In the third section, we study the electron-positron generation and subsequent deflection of positrons by a Gaussian laser beam. The process was investigated by quasi-3D particle-in-cell simulations in the Osiris code \cite{fonseca2002osiris, fonseca2013exploiting}. Both classical and QED descriptions are considered and compared. We examine the effects of radiation reaction on the deflection and on-axis focusing of the positron. The results are compared with the theoretical analysis. We then demonstrate the main benefits of adding an extra plasma channel for positron acceleration to the setup. The fourth section summarizes and discusses the results obtained in this work.

\section{Dynamics of a single positron created within the plane electromagnetic wave}
For on-axis positron guiding during the direct acceleration by the laser pulse, positrons must not escape the most intense central part of the pulse. In realistic experimental conditions, this area will be spatiotemporally limited to a few microns, and such a case will be discussed later in Section \ref{sec3}.
In this section, we theoretically analyze the dynamics of a single positron moving in the plane electromagnetic wave, to obtain an idea about the influence of the initial laser/particle parameters on the positron motion. This model does not fully capture the dynamics of particles in a tightly focused laser pulse, however, it qualitatively demonstrates how the radiation reaction changes the initial positron momentum when it is created in an ultraintense field.
We use the classical approach with radiation reaction, where the exact solution can be obtained, while stochastic QED effects are for now neglected.
We first derive the equations for the positron momentum and then show the positron trajectories for particular values of the initial positron energy and wave field.

\subsection{Positron momentum in a plane electromagnetic wave including classical RR}
In order to investigate positron dynamics, we take the exact solution of the Landau-Lifshitz equation of motion from the classical theory of radiation reaction \cite{landau1987classical}. This is a commonly used approximation of the Lorentz-Abraham-Dirac equation, which is free of nonphysical runaway solutions.
If not explicitly said otherwise, dimensionless units normalized to the wave (laser) frequency $\omega_0$ will be used in this paper, as follows: time $t \rightarrow t\omega_0$; momentum $p\rightarrow p/(m_e c)$; energy $\mathcal{E}\rightarrow \mathcal{E}/(m_e c^2)$; and electric field $E \rightarrow e E/(m_e c \omega_0)$.

In order to establish the positron momentum in our configuration, we first take the general expression of four-velocity for a positron in a linearly-polarized electromagnetic plane wave \cite{piazza2008exact}:

\begin{equation}
\begin{split}
u^{\mu}(\phi)=&\frac{1}{h(\phi)}\Bigl(u_0^{\mu}+\frac{1}{2n^{\beta} u_{0,\beta}}[h^2(\phi)-1]n^{\mu}\\
&+\frac{1}{n^{\beta} u_{0,\beta}}\mathcal{I}(\phi)f^{\mu\nu}u_{0,\nu}+\frac{1}{2n^{\beta} u_{0,\beta}}a^2 \mathcal{I}^2 (\phi) n^{\mu}\Bigr).
\end{split}
\end{equation}

\noindent Here, the Minkowski metric tensor is $\mathrm{diag}(+1, -1, -1, -1)$ and $\beta, \nu$ are summation indices; $u_0^{\mu}$ is the initial positron four-velocity;
$n^{\mu} =(1, \textbf{n})$ with $\textbf{n}$ being a unit vector in the wave propagation direction; $\phi$ is the wave phase; $a$ is the normalized vector potential of the wave, and $f^{\mu\nu}=n^{\mu}a^{\nu}-n^{\nu}a^{\mu}$. The influence of RR is included in the terms:
\begin{equation}\label{eq:h_and_I}
h(\phi) = 1 + h_{\mathrm{RR}}(\phi), \hspace{0.3cm} \mathcal{I}(\phi) = \mathrm{cos} (\phi)h(\phi) -\mathrm{cos}(\phi_0)+\mathcal{I}_{\mathrm{RR}},
\end{equation}
\noindent where
\begin{eqnarray}
h_{\mathrm{RR}}(\phi) &=& \frac{1}{3} \alpha \eta_0 a^2[(\phi-\phi_0)-\mathrm{sin}(\phi)\mathrm{cos}(\phi) +\mathrm{sin}(\phi_0)\mathrm{cos}(\phi_0)], \nonumber \\
\mathcal{I}_{\mathrm{RR}}(\phi)&  =& -\frac{2}{3}\alpha\eta_0(\mathrm{sin}(\phi)-\mathrm{sin}(\phi_0)) \nonumber\\
&&\times [1+\frac{a^2}{3}(\mathrm{sin}^2(\phi)+\mathrm{sin}(\phi)\mathrm{sin}(\phi_0)+\mathrm{sin}^2(\phi_0))].\nonumber\\
\label{eq:h_and_I_RR}
\end{eqnarray}
\noindent Here, $\alpha=1/137$ is the fine-structure constant, $\phi_0$ is the initial phase and $\eta_0 = {\chi_e}_0/a_0$, where ${\chi_e}_0$ is the initial positron quantum parameter and $a_0$ is the magnitude of the normalized vector wave potential. When RR is negligible, ${\chi_e}_0\rightarrow 0$ and $\eta_0 \rightarrow 0$. Hence $ h_{\mathrm{RR}}(\phi) \rightarrow 0$ and  $\mathcal{I}_{\mathrm{RR}}(\phi)\rightarrow 0$ and \eqref{eq:h_and_I} reduces to
\begin{equation}\label{eq:h_and_I_noRR}
h(\phi) = 1, \hspace{0.3cm} \mathcal{I}(\phi) = \mathrm{cos} ( \phi)-\mathrm{cos}( \phi_0).
\end{equation}
Note that the solution in the original work \cite{piazza2008exact} was acquired for electrons, so the formulas here are adjusted for positrons. 

We will now assume the conditions corresponding to the arrangement depicted in Fig. \ref{fig:setup}: the electromagnetic plane wave moves in the $z$ direction and is linearly polarized in the $x$ direction; thus, $n^{\mu} =(1,0,0,1)$, $\phi=t-z$ and $a^{\mu}=(0,a_0,0,0)$. The four-vector potential of such a wave is then $A^{\mu}=(0,A_x,0,0)=(0,a_0\mathrm{cos}(\phi),0,0)$. We assume that the positron initial momentum will be nonzero only in the $x$ direction ${p_x}_0 \neq 0$, while ${p_y}_0={p_z}_0=0$ in the $y$ and $z$ direction, respectively. This is identical to the situation when the electron beam responsible for the pair creation propagates in the $x$-direction (as well as resulting gamma photons from the Compton scattering). In such a case, the newly generated positron would inherit positive momentum in the $x$ direction. We finally get the components of the positron momentum $\textbf{p}=(p_x,p_y,p_z)$:
\begin{align*}
p_x(\phi) &= p_x^{\mathrm{center}}(\phi)+p_x^{\mathrm{osc}}(\phi)= \frac{{p_x}_0}{h(\phi)}- \frac{{a_0 \mathcal{I}(\phi)}}{h(\phi)},\nonumber\\
p_y(\phi) &= 0,  \tageq\label{eq:momentum}  \\
p_z(\phi) &=\frac{1}{h(\phi)}\left(\frac{h^2(\phi)-1}{2\gamma_0}- \frac{{p_x}_0}{\gamma_0} a_0  \mathcal{I}(\phi ) + \frac{1}{2\gamma_0 }a_0^2   \mathcal{I}(\phi )^2\right)  \nonumber.
\end{align*}
\noindent 
Momentum component $p_x$ consists of a centroid term  $p_x^{\mathrm{center}}(\phi)$, which includes the initial momentum ${p_x}_0$, and $p_x^{\mathrm{osc}}$, which accounts for the oscillations in the field of the laser pulse. 
${\chi_e}_0$ is now equal to formula \eqref{chi} with the initial relativistic factor of the positron $\gamma=\gamma_0=\sqrt{1+{p_x}_0^2}$.

\subsection{Effects of positron initial momentum and wave potential on the positron dynamics, dependence on RR}

\begin{figure*}
\centering
\includegraphics[scale=0.97]{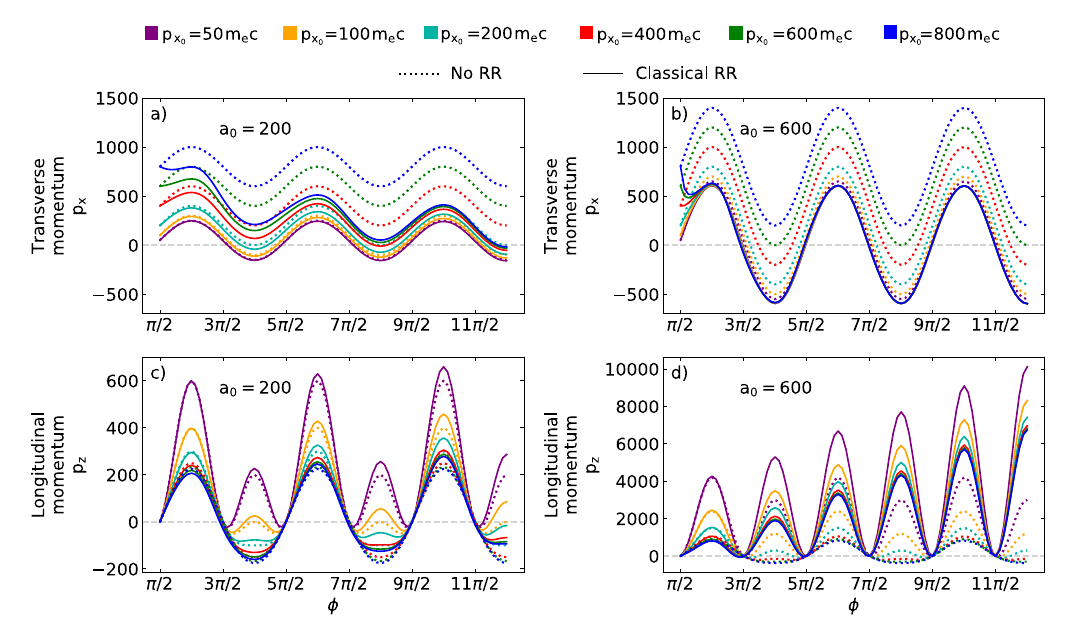}
\caption{Dependence of positron transverse momentum $p_x$ and longitudinal momentum $p_z$ (Eq. \eqref{eq:momentum}) on the wave phase $\phi$: a) transverse momentum $p_x$ for $a_0=200$, b) transverse momentum $p_x$ for $a_0=600$, c) longitudinal momentum $p_z$ for $a_0=200$, d) longitudinal momentum $p_z$ for $a_0=600$. The evolution is shown for different initial transverse momenta ${p_x}_0$. It is represented by dotted lines without RR (Eq. \eqref{eq:h_and_I_noRR}) and by solid lines with classical RR (Eq. \eqref{eq:h_and_I}, \eqref{eq:h_and_I_RR}). The initial phase in every case is $\phi_0=\pi/2$, corresponding to the highest amplitude of $E_x$, where the positrons are created with the highest probability.}\label{fig:theory1}
\label{fig:evolution}
\end{figure*}

\noindent We will now consider various initial conditions and discuss the influence of RR on positron motion.
In Fig. \ref{fig:theory1}, we show the evolution of $p_x$ for a) $a_0=200$ and b) $a_0=600$ and $p_z$ for c) $a_0=200$ and d) $a_0=600$ during the first oscillation cycles $\phi$, as derived in \eqref{eq:momentum}.
The electric field of the wave is pointing in the $x$ direction:
\begin{equation}
E_x=-\frac{\partial A_x}{\partial t}=a_0 \mathrm{sin}(\phi) \frac{\mathrm{\partial} \phi}{\mathrm{\partial}t}= a_0 \mathrm{sin}(\phi).
\end{equation}
The maximum magnitude of the electric field is reached for the multiples of $\pi/2$. These are also the areas of the field, where the creation of electron-positron pairs is most probable. Therefore, without the loss of generality, we consider the initial phase $\phi_0=\pi/2$ in all the studied cases. We varied the values of the initial transverse momentum ${p_x}_0$ from $50~m_e c$ to $800~m_ec$.

For positrons in a focused laser beam, reaching $p_x<0$ after a few oscillation cycles is a necessary condition for the positron deflection. In contrast to the plane wave interaction, positrons that do not reverse the sign of $p_x$ eventually escape the waist of the laser pulse. Therefore, we now look for the trajectory segments where $p_x<0$ is obtained during the positron propagation in the plane wave. This offers preliminary insight into the deflection in a focused Gaussian beam, which will be discussed later in Section \ref{sec3}.

According to Figs. \ref{fig:theory1}a) and b), the amplitude of the oscillations of $p_x$ without RR does not change as they propagate in the wave. For positrons with higher ${p_x}_0$, this means that they never reach $p_x<0$ (${p_x}_0$ from $200$ to $800~m_ec$ for $a_0=200$; and ${p_x}_0$ from $600$ to $800~m_ec$ for $a_0=600$). 
In contrast, when we do not neglect the influence of RR, initial transverse momentum ${p_x}_0$ is effectively attenuated. For all the studied cases, the damping of ${p_x}_0$ happens within the first three oscillation cycles after the creation. However, there is a remarkable difference observed for different values of $a_0$. For $a_0=200$, the decrease of $p_x$ takes more than one cycle for positrons with ${p_x}_0>200~m_e c$. For $a_0=600$, the attenuation is almost immediate, within the very first cycle, regardless of the positron initial momentum. This result indicates that with RR taken into account, a higher $a_0$ assures reaching $p_x<0$ faster and, thus, results in a rapid deflection of positrons in the wave propagation direction.

The influence of RR is also apparent in the evolution of $p_z$. While the~impact is not as distinctive for $a_0=200$, remarkable changes can be observed for $a_0=600$. For instance, for ${p_x}_0=800~m_e c$, without RR, the solution oscillates around $p_z\sim0$, meaning that the positron bounces back and forth in the $z$ direction. With~RR, $p_z$ has positive values and the positron gradually gains energy over time. For ${p_x}_0=50~m_e c$, the positron has positive $p_z$ during the propagation in both cases, with and without~RR; however, with~RR, $p_z$ progressively increases, which is not the case without RR.
Therefore, regardless of the value of ${p_x}_0$, a higher momentum gain can be expected in the direction of the wave propagation direction $z$ when including RR in the calculations.

\begin{figure*}
\centering
\includegraphics[scale=0.97]{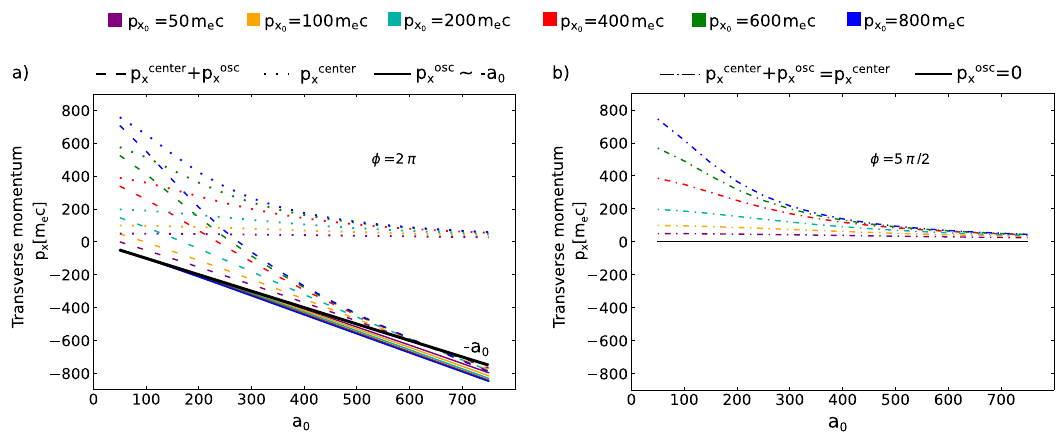}
\caption{Dependence of transverse momentum components with classical RR (Eqs. \eqref{eq:momentum}, \eqref{eq:h_and_I}, \eqref{eq:h_and_I_RR}) on $a_0$ for different phases of the first cycle of the electromagnetic wave: a) $\phi=2\pi$, b) $\phi=5\pi/2$. The initial phase in each case is $\phi_0=\pi/2$. In a), $p_x=p_x^{\mathrm{center}}+p_x^{\mathrm{osc}}$ is depicted with dashed lines; its components $p_x^{\mathrm{center}}$ and $p_x^{\mathrm{osc}}$ are depicted with dotted and solid lines, respectively. The plots are shown for different initial transverse momenta ${p_x}_0$, distinguished by different colors. The values of $p_x^{\mathrm{osc}}$ are approximately equal to $-a_0$; the black line depicts the values of $p_x=-a_0$ for reference. In b), $p_x=p_x^{\mathrm{center}}+p_x^{\mathrm{osc}}=p_x^{\mathrm{center}}$ is depicted with semi-dashed lines for different initial transverse momenta ${p_x}_0$ distinguished by different colors. For all ${p_x}_0$, $p_x^{\mathrm{osc}}=0$ (solid black line).}\label{fig:theory2}
\end{figure*}

We will now study the two components of the transverse momentum $p_x^{\mathrm{center}}$ and $p_x^{\mathrm{osc}}$ as identified in Eq. \eqref{eq:momentum}, for various values of $a_0$ for two different wave phases shown in Fig. \ref{fig:theory2}a) $\phi=2\pi$ and b) $\phi=5\pi/2$. Note that the initial phase is again $\phi_0=\pi/2$. We include RR in all the cases and analyze why its impact increases for larger values of $a_0$. In phase $\phi=2\pi$ (Fig. \ref{fig:theory2}a), $p_x$ takes its minimum value within the first cycle of oscillations. The results show that the influence of the initial positron energy on transverse oscillations becomes less relevant when $a_0$ increases. Also note that for $a_0\geq300$, even particles with ${p_x}_0=800~m_e c$ can revert the initial sign of ${p_x}$ within the first oscillation cycle.

\begin{figure}
\centering
\includegraphics[scale=1]{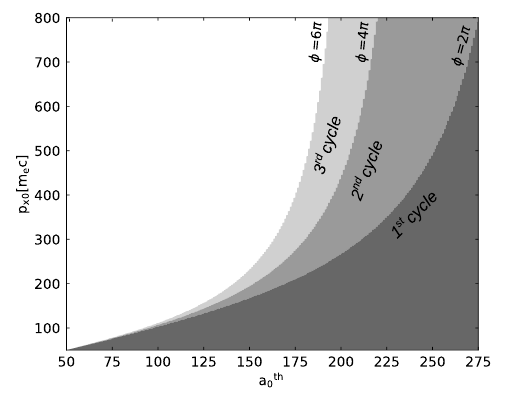}
\caption{The minimum values of $a_0$ required for the deflection in the first three wave cycles: the plot shows the dependence of the initial momentum ${p_x}_0$ on $a_0^{\mathrm{th}}$, the minimum value of $a_0$ necessary for a positron with ${p_x}_0$ to reach $p_x<0$ during a particular oscillation cycle (with classical RR, i.e. Eqs. \eqref{eq:momentum}, \eqref{eq:h_and_I}, \eqref{eq:h_and_I_RR}).
Phases $\phi = 2\pi$, $\phi$ = $4\pi$, $\phi = 6\pi$ correspond to the minimum $p_x$ in the $1^{st}, 2^{nd}, 3^{rd}$ oscillation cycles in the electromagnetic wave, respectively. The areas below each $\phi$ ($\phi = 2\pi$ (dark grey), $\phi$=$4\pi$ (medium grey),  $\phi = 6\pi$ (light grey)) show $a_0$ values for a positron with ${p_x}_0$, where $p_x<0$ is attained during the corresponding cycle. The initial phase in every case is $\phi_0=\pi/2$.} \label{fig:theory3}
\end{figure} 

The fast attenuation of ${p_x}_0$ can be explained when looking at the two terms ${p_x}^{\mathrm{center}}$ and ${p_x}^{\mathrm{osc}}$ in equation \eqref{eq:momentum} separately. The oscillatory part does not change rapidly for different ${p_x}_0$, especially for lower $a_0$.
The significant change is visible in the $p^{\mathrm{center}}$ formula. For high $a_0$, $h(\phi)$ attenuates $p^{\mathrm{center}}$ to almost zero within the first cycle even for the highest ${p_x}_0$ studied here. To put it simply, the higher ${p_x}_0$, the higher $h(\phi)$, and this effect increases with $a_0$. This behavior induces $p_x$ oscillations with amplitude $\sim a_0$ around $p_x\sim0$ for high values of $a_0$, as we already observed in Fig. \ref{fig:theory1}b). 

Phase $\phi=5\pi/2$, depicted in Fig. \ref{fig:theory2}b), corresponds to the endpoint of the first $p_x$ oscillation cycle. The term $p^{\mathrm{center}}$ decreases with higher $a_0$ and ${p_x}^{\mathrm{osc}}=0$. As a consequence, $p_x$ approaches zero in limit with increasing $a_0$.
This observation confirms the established idea that strong RR attenuates the transverse momentum of a particle, therefore, this enables oscillations around $p_x\sim0$.

These results indicate that for each ${p_x}_0$, there is a threshold normalized vector potential $a_0^{\mathrm{th}}$ which assures reaching ${p_x}<0$ in a particular oscillation cycle. Such a map of $a_0^{\mathrm{th}}$ for the 3 first cycles is depicted in Fig. \ref{fig:theory3}. We assign to $a_0^{\mathrm{th}}$ the minimum value of $a_0$ necessary to reach $p_x<0$ during a particular oscillation cycle. 
As shown in Fig. \ref{fig:theory1}, the minimum $p_x$ in each cycle is reached for $\phi$ equal to even multiplies of $\pi$. So when the values at these points are negative, it is guaranteed that $p_x<0$ is reached during the corresponding cycle. The image shows that without sufficient intensity, highly energetic particles do not experience $p_x<0$ during the first few oscillations at all. For instance, for $a_0^{\mathrm{th}}\lesssim190$, $p_x$ does not reverse sign for a positron with ${p_x}_0=800$ during the first 3 cycles.
Also, we can see that for $p_x \lesssim 100~m_ec$, negative momentum is reached within the first cycle when $a_0 \gtrsim {p_x}_0$, or it is not achieved during the first 3 cycles at all. This is in accordance with the previous observation in Fig. \ref{fig:theory1}a), b), where the effect of RR on $p_x$ is less significant for low-energy positrons.

We will now analyze the trajectories of positrons during the first two oscillation cycles with and without RR for $a_0=200$ and $a_0=600$. The values of coordinates $x$ and $z$ are obtained by numerical integration implemented in Python $scipy$ library, which uses a technique from the Fortran library \textit{QUADPACK}. Using $p_x$ and $p_z$ from Eq. \eqref{eq:momentum}, we calculated $x = \int_0^t p_x (t)/\gamma(t)\mathrm{d}t=\int_{\phi_0}^{\phi}
 p_x (\phi )/(\gamma(\phi ) - p_z (\phi ))d\phi$ and 
$z = \int_0^t p_z (t)/\gamma(t)\mathrm{d}t=\int_{\phi_0}^{\phi} p_z (\phi )/(\gamma(\phi ) - p_z (\phi ))d\phi$.
Here, $\gamma(\phi ) - p_z(\phi )=\gamma_0- {p_z}_0=\sqrt{1+{{p_x}_0}^2}$ is the integral of motion.

\begin{figure*}
\centering
\includegraphics[scale=0.84]{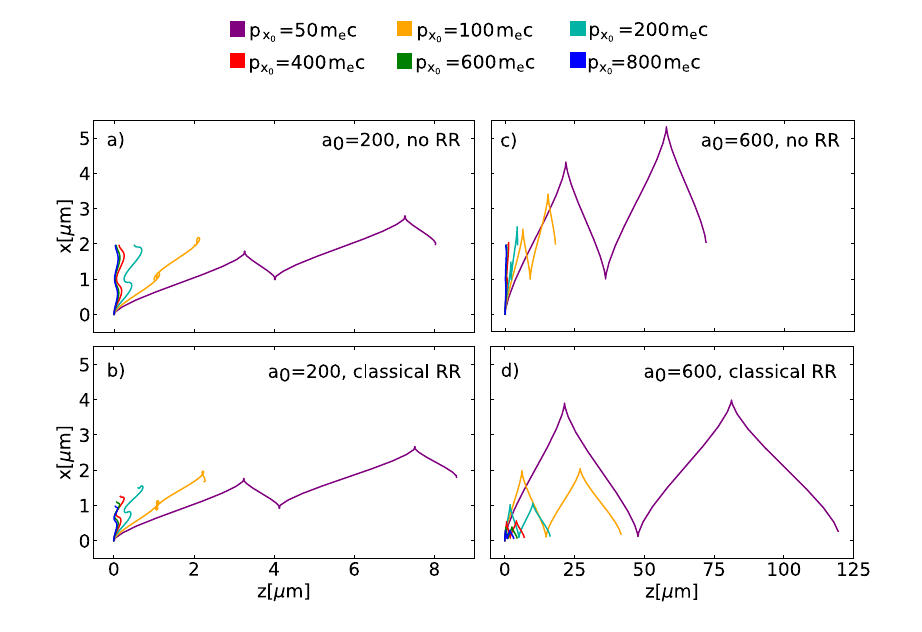}
\caption{Trajectories of positrons between the $\phi = \pi/2$ and $\phi = 9\pi/2$: a) for $a_0=200$ without accounting for RR, b)  for $a_0=200$ with RR, c)  for $a_0=600$ without RR, d)  for $a_0=600$ with RR.
The evolution is shown for different initial transverse momenta ${p_x}_0$. The values of $z$ (wave propagation direction) and $x$ (wave polarization direction) are shown in microns to help assess which positrons may leave the laser focus in the transverse direction $x$.}\label{fig:theory1_2}
\end{figure*}

The results are depicted in Fig. \ref{fig:theory1_2}. As expected, the effect of RR on particle trajectories is more significant for $a_0=600$. Moreover, for $a_0=200$, the transverse motion is more notable compared to the forward drift for high values of ${p_x}_0$. While RR slightly suppresses this effect for these highly energetic positrons $({p_x}_0\geq 200~m_ec)$, these positrons still propagate through longer distances in the $x$ direction than in the $z$ direction. In contrast, for $a_0=600$, the amplitude of oscillation is essentially suppressed. In particular, for ${p_x}_0>200~m_ec$ the amplitude is less than $x=1~\mathrm{\upmu m}$. Also, note that the oscillation amplitude increases with decreasing ${p_x}_0$. Thus, if the initial transverse momentum is small, the high-amplitude oscillation might, in principle, cause undesirable escape from a very focused laser spot size in realistic conditions.

The main outcomes of this theoretical model are that 1) the initial transverse momentum of positrons gained during the Breit-Wheeler process is suppressed during the first oscillation cycles in the electromagnetic wave and 2) this suppression is more rapid with increasing intensity.
To conclude the consequences of the analysis of the plane wave, positrons are likely to escape the~finite spot size of a realistic laser pulse if the laser intensity is low and, consequently, the radiation reaction is not effective in this case. This is shown with particle-in-cell simulations in the next section, where a laser pulse with a finite spot size is used.

\section{Deflection of positrons by a focused Gaussian laser beam}\label{sec3}
\noindent In this part, we will assume a~Gaussian laser beam instead of the plane wave. The study will be carried out via particle-in-cell simulations. We simulate the whole process including positron creation and deflection and discuss the impact~of~RR on the positron deflection towards the laser axis. We also compare the classical approximation with the QED~approach which is more accurate for the conditions studied here. Furthermore, the benefits of adding a plasma channel into this setup are demonstrated. Here, by the \textit{deflection of positrons}, we mean their deflection in the \textit{direction of the laser pulse propagation}, i.e. maintaining them within the area of the laser pulse diameter for at least 190~fs after the creation started.

\subsection{Simulation parameters}
\label{subsec:IIIA}
\noindent The scenario with the Gaussian laser beam is simulated with the particle-in-cell framework OSIRIS \cite{fonseca2002osiris,fonseca2013exploiting}. We employ a quasi-3D algorithm, where the fields and currents are expanded into azimuthal Fourier harmonics (modes), while macroparticles move in the 3D~laser dynamics \cite{davidson2015implementation}. The linearly polarized laser can be modeled with the first harmonic \cite{lifschitz2009particle}. Hence, only two modes are sufficient to describe both the fields and currents that are radially symmetric ($z,r$ coordinates) and the laser field  ($z,x$ coordinates). Consequently, the quasi-3D approach benefits from significantly less demanding computational requirements than the 3D~geometry but also assures a more accurate approximation of the 3D~space than 2D~simulations, where the field falls off as $\sim 1/r^2$.

The laser pulse propagates along the $z$ axis and is polarized in the $x$ direction, with a~Gaussian spatial field profile. It has the maximum normalized vector potential $a_0$ (values are specified later), wavelength $\lambda_0 = 1~\mathrm{\upmu m}$, FWHM temporal duration $\tau_L=150$~fs and waist $w_0 = 3.2~\mathrm{\upmu m}$. 
The focal plane is located at the position of $z = 246.7~\mathrm{\upmu m}$. In this paper, we will refer to the laser pulse diameter as $d_L=2\sqrt{2 \mathrm{ln}(2)}w_0\sim 7.5~\mathrm{\upmu m}$ at full width at the quarter maximum of laser electric field $a_0/4$. Within this area, 90\% of the electric field distribution $E_x$ and 98\% of the intensity distribution is contained.

With the quasi-3D geometry, we cannot represent the electron beam at 90 degrees properly, but we can represent the photons. The center of the gamma beam is near the laser focus at $z=242.7~\mathrm{\upmu m}$. The spatial profile in the $z$ direction is Gaussian with~FWHM~of~$3.2~\mathrm{\upmu m}$ and the radial distribution is uniform. The beam has a~synchrotron energy distribution, calculated according to Ref. \cite{erber1966high}, and a density of $1.1\times10^{17}~\mathrm{cm}^{-3}$. This value corresponds to a lower bound estimate of gamma rays generated by the collision of the laser pulse with a$~\sim 2$~GeV electron beam of 10~pC of charge occupying a volume of $10 \times 1 \times 1 ~\mathrm{\upmu m^3}$, which was derived as follows. We assumed that only 10\% of the 10~pC beam will radiate due to the geometry of interaction (based on previous results from full-3D simulations). As a result, there are, at least, $6 \times 10^6$ gamma rays created in the beam area. This corresponds to a density of $2 \times 10^{17} /\mathrm{cm^3}$, and we take $10^{17} /\mathrm{cm^3}$ as a conservative lower bound estimate. More details on the calculation and the spectrum profile can be found in the Supplementary Material of our previous work\cite{martinez2023creation}.  We chose a 2 GeV electron beam as it is within the reach of various facilities\cite{wang2013quasi,shin2018quasi,gonsalves2019petawatt} and also as we proved that boosting it to 10-20 GeV would only increase the number of positrons by a factor of order unity\cite{martinez2023creation}.

\begin{figure}
\centering
\includegraphics[scale=0.5]{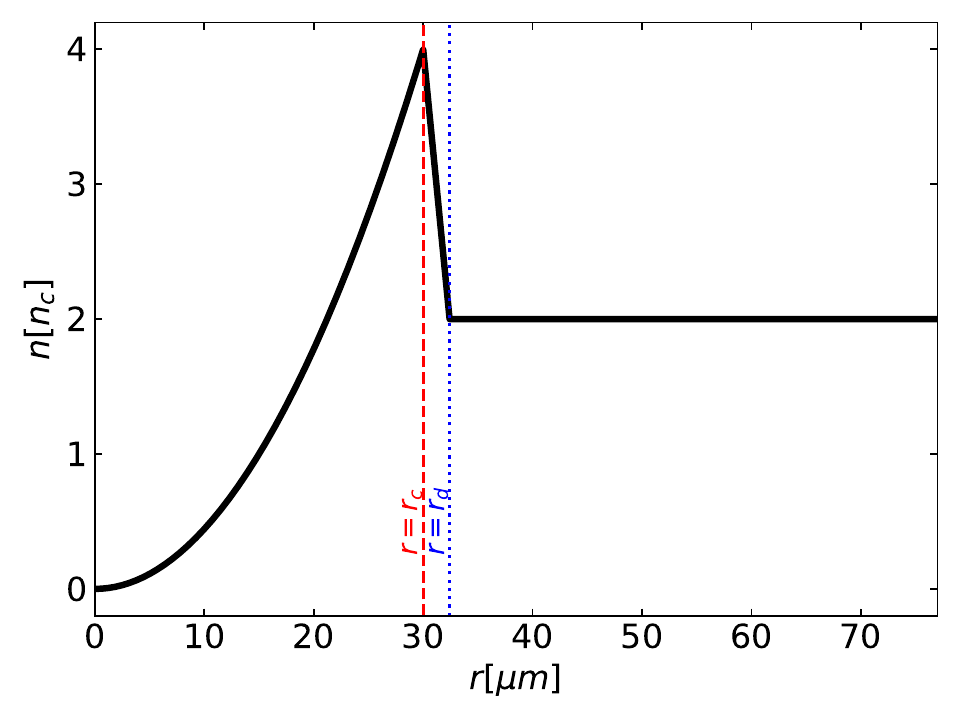}
\caption{Radial density profile $n$ of the plasma channel in the simulations for $z>z_1= 211.7~\mathrm{\upmu m}$, according to formula \ref{eq:density} (black solid line). The density profile has a parabolic increase from $r=0$ up to channel radius $r=r_c=30~\mathrm{\upmu m}$ (red dashed line), reaching the peak of $n=4n_c$. The density is then rapidly linearly dropping until $r=r_d=32.4~\mathrm{\upmu m}$ (blue dotted line). For $r>r_d$, it remains constant as $n=2n_c$, until the simulation box boundary at $r=L_r=77.7~\mathrm{\upmu m}$.} \label{fig:simulation0}
\end{figure}

In general, we performed two kinds of simulations: 1) in a vacuum, where no particles with the exception of gamma photons are initiated at the beginning, and 2) with the plasma channel filled with fully ionized nitrogen plasmas of 2~eV temperature. In the plasma channel case, for $z>z_1= 211.7~\mathrm{\upmu m}$, we introduced the channel radial profile as 
\begin{equation}\label{eq:density}
 n(r)=\begin{cases}
   \left(10^{-3}+(4-10^{-3})\left(\frac{r}{r_c}\right)^2\right)n_c & \text{if $r\leq r_c$},\\
   \left(4-\frac{2}{r_d-r_c}(r-r_c)\right)n_c  & \text{if $r>r_c$ \& $r\leq r_d$},\\
     2 n_c & \text{if $r>r_d$}.
  \end{cases}
\end{equation}
The plasma channel radius is $r_c=30~\mathrm{\upmu m}$ and $n_c=m_e \varepsilon_0 \omega_0^2/e^2=1.1 \times 10^{21}~\mathrm{cm^{-3}}$ is the critical density, where $\varepsilon_0$ is the vacuum permittivity. For $r>r_c$, the density linearly drops until $r=r_d=32.4~\mathrm{\upmu m}$ and then plateaus for $r>r_d$. This radial density profile is depicted in Fig. \ref{fig:simulation0}. For $z<z_0=206.1~\mathrm{\upmu m}$, we put density $n=0$. For the range between $z\geq z_0~\mathrm{\upmu m}$ and $z\leq z_1$, we multiplied Eq. \ref{eq:density} by a factor of $(z-z_0)/L$. This introduces a short linear density gradient of $L=5.6~\mathrm{\upmu m}$, to avoid a step-like density entrance into the plasma channel.

The size of the simulation domain is $L_z=203.7~\mathrm{\upmu m}$ and $L_r= 77.7~\mathrm{\upmu m}$. The size of one cell is $\Delta z = 15.9~\mathrm{nm}$ and $\Delta r = 15.9~\mathrm{nm}$, in the $z$ and $r$ direction, respectively. We use 16 macroparticles per cell for leptons and ions, and 128 macroparticles per cell for gamma photons. The equations of motion were solved by the Boris pusher\cite{boris1970relativistic}. The duration of one timestep is $\Delta t=27~\mathrm{as}$. We also performed convergence tests, because simulations with ultrahigh intensities can lead to numerical  inaccuracies \cite{arefiev2015temporal, vranic2016classical, gordon2017pushing, robinson2019extreme, tangtartharakul2021particle, li2021accurately} if the timestep is not adequate. The tests are included in Appendix.

\subsection{Effects of RR on the dynamics of positrons generated in the focus of a Gaussian laser beam}

\begin{figure*}
\centering
\includegraphics[scale=1]{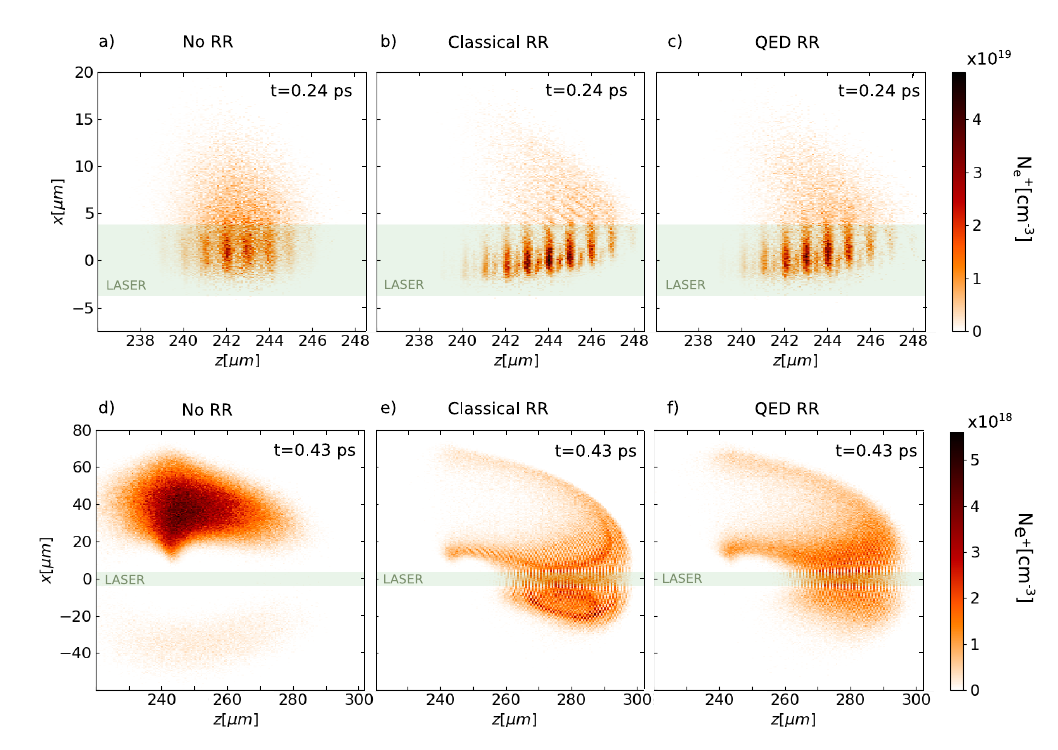}
\caption{Number of positrons $N_{e^{+}}$ per $\mathrm{cm^{3}}$ in the range of $-0.05~\mathrm{\upmu m} <y <0.05~\mathrm{\upmu m}$, where the laser pulse has the highest intensity, is shown in the 2D plane $z-x$. Cases with a) no RR, b) classical RR, and c) QED RR are shown for simulation time $t=0.24~\mathrm{ps}$, and d) no RR, e) classical RR, and f) QED RR for simulation time $t= 0.43~\mathrm{ps}$. The light green areas labeled as "LASER" correspond to the area of $d_L$, where the positrons are considered deflected.} \label{fig:simulation1}
\end{figure*}

We first examine positron creation and deflection in the vacuum. The laser pulse of $a_0=600$ is taken as a case study. For our choice of parameters, this corresponds to the intensity of $4.9\times10^{23}~\mathrm{W/cm^2}$, the power of~$\sim80~\mathrm{PW}$ and approximately $12$~kJ of energy. The simulation duration is 800~fs. We compare the three following simulation cases: 1) without RR; 2) with the classical RR module; 3) with the QED~RR module. The creation of positrons in all the cases started at $t=0.21~\mathrm{ps}$. The comparison of positron density is shown at two simulation times: at $t=0.24$~ps (shortly after the creation starts), and $t=0.43$~ps (after the creation finishes), in Fig. \ref{fig:simulation1}a)-c) and d)-f), respectively. In the rest of this section, time and length are expressed in SI units, and other quantities are expressed in dimensionless units used in the previous section.

According to the simulation results, there is already a slight difference in the positron density distribution shortly after their creation (Fig. \ref{fig:simulation1}a)-c)) influenced by RR. However, without RR, most of the positrons are sustained in the area of the laser focus in a similar manner as with RR. In contrast, there is an obvious difference between the deflection with RR and when RR is neglected, for both the classical and QED approaches (Fig. \ref{fig:simulation1}d)-f)). Without RR, none of the positrons created are sustained on the laser axis and deflected. This is in contrast with the RR cases, where a significant fraction of the positrons can be preserved within the pulse diameter and oscillate in the laser field. These results confirm the previous assumption that RR is responsible for focusing positrons on the axis.

In general, we can observe certain dissimilarities in the positron distribution between the classical and QED regimes, although the results agree qualitatively well. Differences between the classical and QED approximations have been already observed in the simulations of the orthogonal geometry previously \cite{geng2019quantum}, where the QED probabilistic nature permitted some behavior that is not allowed in the classical regime. In our simulations, the quantum parameter of newly created positrons reaches values up to ${\chi_e}_0\sim5$, which corresponds to the QED-dominated regime of interaction. Consequently, some differences are observed between the classical and QED simulation approaches. Approximately 7\% more positrons are created in the QED case due to the stochastic behavior ($1.70 \times 10^7$ vs. $1.83 \times 10^7$ positrons). However, there is a lower deflection rate in the QED case. We observed that $\sim$13\% fewer positrons are deflected in the QED regime compared to the classical RR regime ($1.6 \times 10^6$ vs. $1.4 \times 10^6$ positrons). This difference is relatively minor; therefore, the classical regime, also used in the theoretical analysis above, can bring acceptable qualitative estimates of the positron dynamics when the laser has a high field amplitude of $a_0=600$.

\begin{figure*}
\centering
\includegraphics[scale=1]{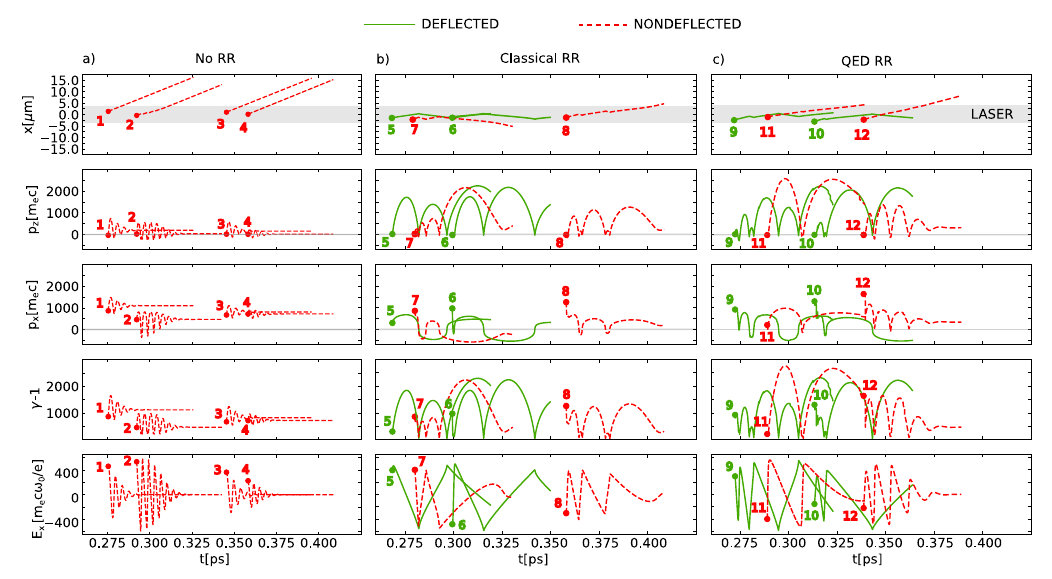}
\caption{Examples of dependence of $x$, $p_z$, $p_x$, $\gamma$-1 (kinetic energy in $m_e c^2$), and $E_x$ experienced by the positron at simulation time $t$. The test positron macroparticles are shown for simulations with a) no RR, b) classical RR and c) QED RR. Dashed red lines correspond to the positrons that were not deflected (1-4, 7, 8, 11 and 12); solid green lines correspond to the positrons that were deflected in the direction of the pulse propagation (5, 6, 9 and 10). The evolution for each positron is shown for the first 50~fs of propagation after its creation. The dots indicate the spots of their creation. The light grey rectangles correspond to the area of laser diameter $d_L$, where the positrons were deflected in the direction of the laser pulse. The grey lines for $p_z=0$ and $p_x=0$ are shown for better visualization.} \label{fig:simulation2}
\end{figure*}

We now study examples of dynamics for 4 test positrons for each of the simulation cases. In Fig. \ref{fig:simulation2}a), the positrons are not influenced by the RR. All the positrons escaped the area of the laser diameter $d_L$ in the first few cycles. After the escape, they did not gain additional $p_z$ and $p_x$ boost. Their maximum extra energy gain in addition to the creation energy is always given by the amplitude of their last oscillation in the laser field. Particles no. 1, 3, 4 that experience initial laser field $E_x = {E_x}_0\lesssim 500$ do not even reach $p_x<0$ while oscillating in the laser field. Particle no. $2$ is created at $E_x  = {E_x}_0\sim a_0 $ and its $p_x$ decreases slightly below zero during the first three oscillations. However, this is not sufficient, since ${p_x}^{\mathrm{center}} > 0$ and $|{p_x}^{\mathrm{center}}|\sim |{p_x}^{\mathrm{osc}}|$. As a consequence, the particle escapes $d_L$ after 6~oscillation cycles. This matches well with the theoretical prediction in Fig. \ref{fig:theory1}b. 

In comparison, with the classical~RR (Fig. \ref{fig:simulation2}b), some positrons can be effectively sustained on~the~axis. For particles no. 5 and 6, the suppression of the initial momentum occurs within the first cycle, after which $p_x$ oscillates around ${p_x}^{\mathrm{center}}\sim 0$. This enables preserving of the particle in the highest-value laser field, and the amplitude of $p_z$ and $\gamma$ is correspondingly high. Note that for both these particles, the initial laser field has values ${E_x} = {E_x}_0 \gtrsim 400$, which ensures a sufficient impact of RR on the dynamics. Opposed to positrons no. 5 and 6, positrons no. 7 and 8 are not deflected. For particle no. 7, the transverse momentum oscillates around ${p_x}^{\mathrm{center}}\sim 0$, and it is created in  ${E_x} = {E_x}_0 \gtrsim 400$, close to $a_0$. However, the initial transverse position $x = x_0$ is too far from $x=0$. Consequently, the positron is not maintained in the area of the laser pulse and escaped in the negative $x$ direction, not completing the second oscillation cycle. Particle no. 8 is created in a lower field (${E_x} = {E_x}_0  \sim 300$) and RR does not provide a decrease to $|{p_x}^{\mathrm{center}}|\sim 0$, which resulted in the escape in the positive $x$ direction.
 
Outcomes from the QED simulation (Fig. \ref{fig:simulation2}c) are comparable with the classical case. However, we can distinguish the stochastic suppression of $p_x$ for deflected positrons (no. 9 and 10). Positron no. 9 loses most of its initial momentum in the second cycle, while it is propagating toward $x=0$, where the energy loss is more probable. Positron no. 10 suddenly loses a big portion of the energy in a step-like manner. This also happens as the positron propagates closer towards the laser axis $x=0$ with the highest value of $E_x$. In contrast, positron no. 11 escapes the pulse diameter with almost no ${p_x}$ loss. This behavior is only possible due to the stochastic effects, similar to quantum quenching \cite{harvey2017quantum}. Particle no.~12 escapes because $|{p_x}^{\mathrm{center}}|>>0$ (similarly to no. 8).

\begin{figure}
\centering
\includegraphics[scale=1]{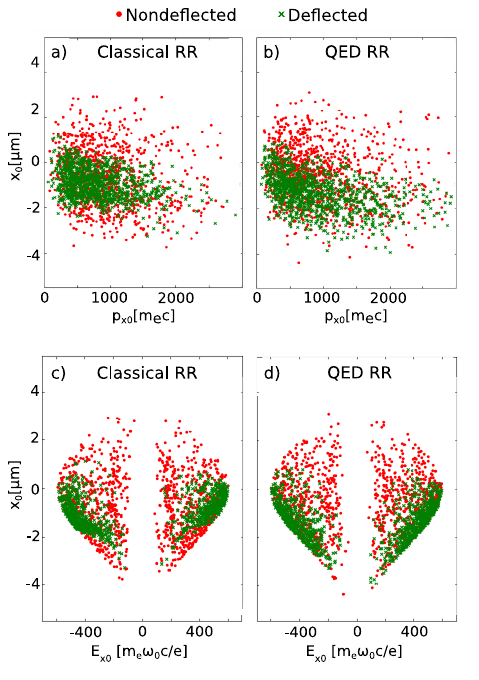}
\caption{The initial transverse phasespace of deflected (green) and non-deflected (red) particles. Here, $x_0$ is the initial position, and  ${p_x}_0$ is the initial transverse momentum for a) classical RR and b) QED RR; Initial distribution of the same particles according to $x_0$ and the transverse electric field ${E_x}_0$ experienced by the positrons right after their creation for c) classical RR and d) QED RR. The representative positron macroparticles are randomly selected. The corresponding simulation time was $t=0.43~\mathrm{ps}$.} \label{fig:simulation3}
\end{figure}

These results show that the initial conditions play a significant role in the subsequent motion of positrons, since the whole deflection/escape process happens quickly, during a few oscillation cycles. 
We will now take a look at the difference in initial conditions between positrons which were deflected and nondeflected. In Fig. \ref{fig:simulation3}a), b) we plot the initial momenta ${p_x}_0$ of randomly chosen positron macroparticles and their corresponding initial $x$ coordinates $x_0$, for a) classical approximation and b) QED approximation. The initial momentum ranges up to almost $3000~m_e c$. The images show that the deflection is allowed for any initial momentum, as already hinted by the analytical theory. 
However, there is an apparently higher probability of deflection for the positrons that were produced early, in the area of the laser pulse where $x_0\leq0$. 

This can be explained by Fig. \ref{fig:simulation3}c), d), where the dependence of ${E_x}_0$ - the value of the laser field positrons experience right after the creation - is shown for classical and QED cases, respectively. 
In general, positrons created at $x_0>0$ immediately appear in the area of lower intensity with an initial momentum "outwards" (away from the centre of the pulse). These positrons do not cross the laser axis $x=0$ to experience the highest laser field, which makes their deflection less probable. In contrast, positrons with $x_0<0$, still have, in principle, a chance to get to the central part of the laser pulse, where they can effectively deplete their momentum (as no. 10 in Fig. \ref{fig:simulation2}).

Moreover, the lower $x_0$ is, the lower ${E_x}_0$ can be for the positron to get deflected. This is due to the fact that positrons with low values of $x_0$ have more interaction time available to get to the central part of the laser beam, where their initial momentum can be depleted more effectively. Note that for classical RR (Fig. \ref{fig:simulation3}c), the positrons below $x_0<-2$ are more rarely deflected than in the QED case (Fig. \ref{fig:simulation3}d). This is due to the fact that with classical RR, these positrons escape transversely in the direction of negative $x$ (as no. 7 in Fig. \ref{fig:simulation2}). In contrast, the stochastic nature of QED RR allows the positron to reach the center of the laser beam due to a less significant momentum loss. As a consequence, the oscillation center is shifted towards the positive $x$ direction, thus, the positrons oscillate near $x\sim0$, and their escape in the negative $x$ direction is less likely.

\subsection{Dependence of the deflection on the laser intensity}
\begin{figure*}
\centering
\includegraphics[scale=1]{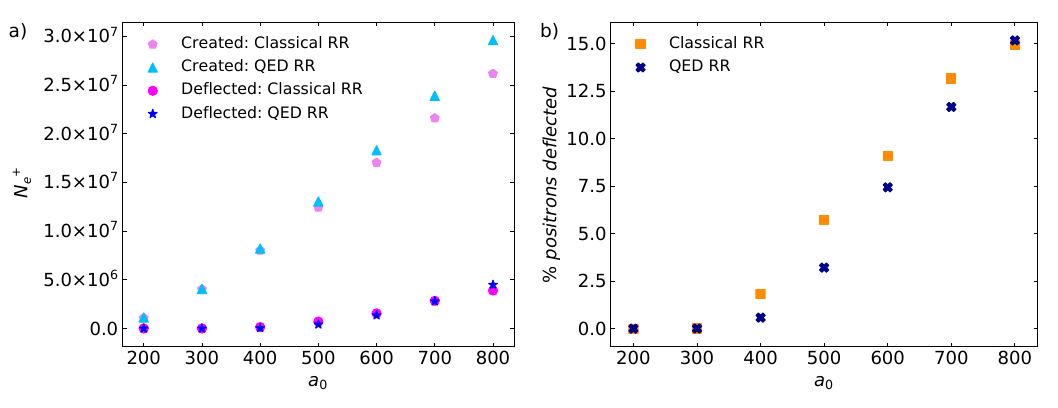}
\caption{Number of positrons $N_{e^+}$, created and deflected as a function of $a_0$: a) absolute numbers of positrons: created with classical RR (light pink pentagon); created with QED RR (light blue triangle); deflected with classical RR (magenta circle); deflected with QED RR (blue star),
and b) ratio of positrons deflected/created in percentage: classical RR (orange square); QED RR (dark blue cross).
The positrons labeled as "created" are all the positrons that were in the simulation window at $t=0.43~\mathrm{ps}$; "deflected" positrons are the ones that were at the area of the laser diameter $d_L$ at $t=0.43~\mathrm{ps}$.
}\label{fig:simulation5}
\end{figure*}

\noindent All the results presented so far indicate that the intensity of the laser pulse plays a crucial role in the positron deflection. It implies that $a_0$ chosen for the laser pulse in the experiment can directly influence the final positron rates. In Fig. \ref{fig:simulation5}a), we show the number of positrons created and deflected for different $a_0$. Not only the number of positrons rapidly increases for high $a_0$, but we can also see that the deflection does not occur at all for $a_0=200$ and $a_0=300$, despite a certain number of positrons created.
In Fig. \ref{fig:simulation5}b), the ratio of positrons created/deflected is depicted. For $>1\%$ deflection of positrons in the QED case, at least $a_0=400$ is required. For our parameters, this means that at least a 35 PW laser system would be necessary to achieve such a deflection and make subsequent direct acceleration possible.

\subsection{Additional direct laser acceleration of positrons in a plasma channel}

We will now investigate the effects of the plasma channel presence in the setup. In the plasma channel, the laser pulse can be effectively guided for several hundreds of microns. Therefore, the time of propagation of positrons inside the laser focus can be prolonged. The presence of plasma can also induce the DLA process because the frequency of positron betatron oscillations can reach values close to the Doppler-shifted laser frequency. This induces resonant coupling between the positron oscillations in the laser field and the betatron oscillations. The focusing of positrons is provided by beam loading of a copious amount of electrons on the propagation axis \cite{martinez2023creation}.

\begin{figure*}
\centering
\includegraphics[scale=1]{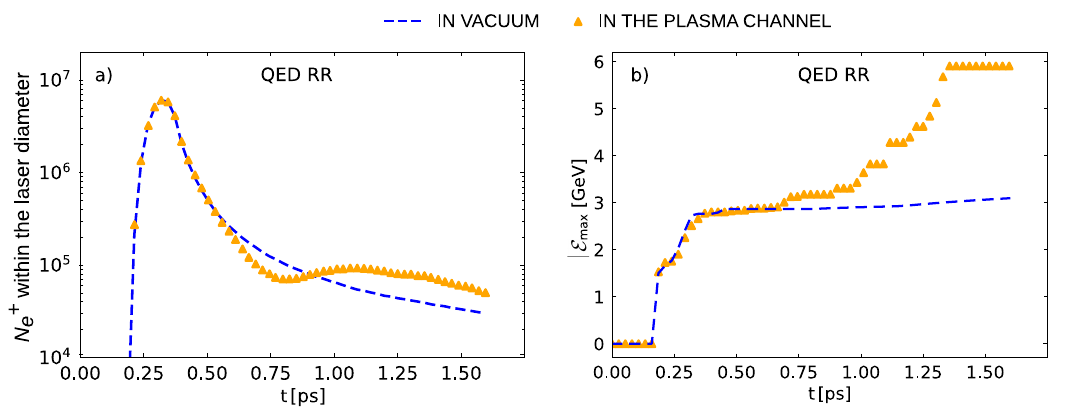}
\caption{a) The number of positrons ${N_e}^+$ sustained inside the area of the laser pulse diameter $d_L$, b) maximum positron energy $\mathcal{E}_{max}$ reached in vacuum (blue dashed line) and plasma (orange triangles), as a function of simulation time $t$. The laser has $a_0=600$. The simulations in this figure were performed with the quantum model of RR (QED RR).
}\label{fig:simulation6}
\end{figure*}

For simplicity, we assume here that the whole process, including pair generation, deflection, and acceleration, takes place in the plasma. In practice, some space between the collision and plasma channel entrance would be present. We compare the temporal evolution of the number of positrons within the area of the laser pulse diameter $d_L$ for the vacuum case and the plasma case (see Fig. \ref{fig:simulation6}a). The simulation duration is $1.6~\mathrm{ps}$. 
The creation and deflection are not affected by the plasma presence. However, after $t=0.43~\mathrm{ps}$, the number of positrons within the laser pulse diameter varies. Due to effective electron beam loading during the DLA process, more positrons can be focused on the axis. In principle, if the number of positrons generated is close to the number of electrons beam loaded, it could possibly lead to retaining a bigger fraction of the positrons on the axis.

Moreover, the maximum final energy obtained by the positrons in the plasma can be doubled compared to the vacuum case (see Fig. \ref{fig:simulation6}b). This gradual increase in energy by DLA starts shortly after the deflection, and it effectively lasts for $\sim0.7~\mathrm{ps}$.  
As a consequence, the creation and acceleration of positrons up to $\sim6$~GeV in approximately half-a-mm distance is possible.
For a more detailed study of the acceleration stage see Ref. \cite{martinez2023creation}.

\section{Summary and discussion}
 The orthogonal collision of a relativistic electron beam with a~multi-PW laser pulse enables the creation of $e^{-}e^{+}$ pairs by the nonlinear Breit-Wheeler process. A fraction of the positrons can be deflected in the direction of the laser pulse propagation and become available for direct laser acceleration in a plasma channel. This process is crucially dependent on the radiation reaction recoil, naturally acting on relativistic particles moving in high-intensity fields. In recent studies \cite{ji2014radiation,vranic2018extremely}, it has been shown that radiation reaction causes on-axis radiative trapping of particles from a target, by compensating the expelling ponderomotive force. These particles could have been assumed as stationary, with a nonzero initial transverse momentum. In contrast, here, where the positrons were created already with a certain energy at the most intense laser part, the role of RR was to attenuate this initial transverse momentum, so they do not escape the laser axis in the first place.
 
 A realistic laser pulse is focused into a few-micron radius. Therefore, in order to deflect positrons in the direction of its propagation, the transverse escape of positrons from the finite laser spot size needs to be avoided. This means that their initial transverse momentum ${p_x}_0$, gained during the pair creation, has to be damped during the first few oscillation cycles inside the laser field. This is provided by the radiation reaction, responsible for guiding positrons on the laser axis. For positrons with initial transverse momentum ${p_x}_0 > 100~m_e c$, the decrease of transverse momentum is more dependent on the laser intensity rather than ${p_x}_0$. Consequently, the positron deflection depends mostly on the field amplitude $a_0$, and only has a weak dependency on the initial momentum itself. Very high fields of $a_0 \geq 400$ are necessary for $>1$\% deflection of positrons in the direction of the laser pulse propagation. For $a_0 \leq 300$, we did not observe any deflection despite the fact that $\sim10^5$ positrons were created. These results, obtained by the particle-in-cell simulations, compared for both classical and QED approximation, are well supported by the theory based on the classical Landau-Lifshitz equation of motion.
 In addition, a plasma channel can be placed after the location of the collision, enabling subsequent direct laser acceleration of positrons. This additional acceleration  can double the maximum final positron energy, reaching several~GeV, and almost double the number of positrons in the beam. 
 
For our example of $a_0=600$, about $100~\mathrm{fC}$ can be deflected immediately after the creation. Despite this fact, the number of positrons in the laser focus gradually decreases with time, since the positrons escape as they propagate further through vacuum or plasma. The charge of the beam (within $d_L$) in the simulation was on the order of $\sim\mathrm{fC}$ at the time of $1.4$~ps after the creation started. One of the natural ways to increase the final charge would be to increase the laser intensity. However, although the rapid progress in laser development, such intensity is currently out of reach. There are nevertheless other alternatives of boosting the final positron charge. The first path to enhance the positron number is to increase the charge of the incident electron beam. The electron energy of 2 GeV is available in many laboratories worldwide using LWFA. The charge of these beams is about 10-100 pC \cite{wang2013quasi,shin2018quasi,gonsalves2019petawatt}. Increasing the charge of the incident electron beam by a factor of 10 from 10 pC to 100 pC would increase the number of positrons by a factor of ten as well, from 0.1 pC to 1 pC. Moreover, the examination of another angle of interaction than $90^{\circ}$ could bring more insight into the process, possibly showing more promising geometry for positron deflection. Another option could be to generate positrons by the Bethe-Heitler process, by impinging the laser beam on a solid target instead of a relativistic $e^{-}$ beam, and place the plasma channel behind the target. This method might presumably increase the charge while lowering the requirements on the laser power. This will be investigated in our future work.

\begin{acknowledgments}
This work is supported by IPP - Mobility II: CZ.02.2.69/0.0/0.0/18\_053/0016925; European Regional Development Fund-Project “Center for Advanced Applied Science” (No. CZ.02.1.01/0.0/0.0/16\_019/ 0000778); Ministry of Education, Youth and Sports of the Czech Republic through the e-INFRA CZ (ID:90140); Grant Agency of the Czech Technical University in Prague, grant no. SGS22/185/OHK4/3T/14; European Research Council (ERC-2015-AdG Grant No. 695088); Portuguese Science Foundation (FCT) Grant No. CEECIND/01906/2018 and PTDC/FIS-PLA/3800/2021. We acknowledge PRACE for access to MareNostrum based in the Barcelona Supercomputing Centre. Additional computational resources were provided by the Oblivion supercomputer in The High Performance Computing Center at the University of Évora, Portugal.
\end{acknowledgments}

\appendix

\section*{Appendix: Numerical convergence of the simulations} \label{sec:Appendix}

\begin{figure}

\centering{
\includegraphics[scale=0.5]{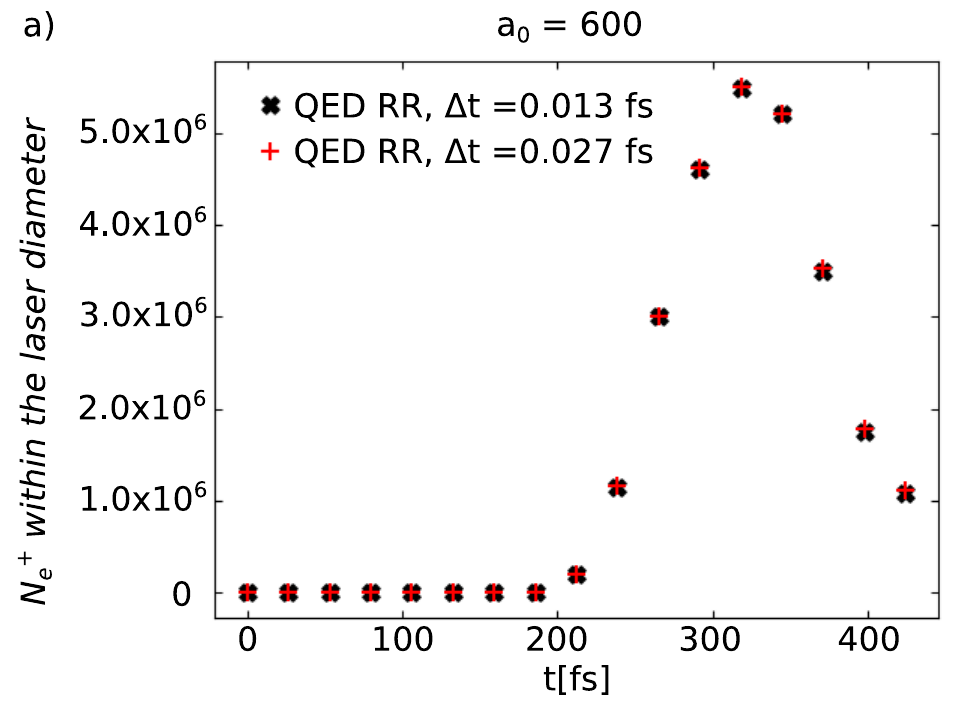}}

\centering{
\includegraphics[scale=0.5]{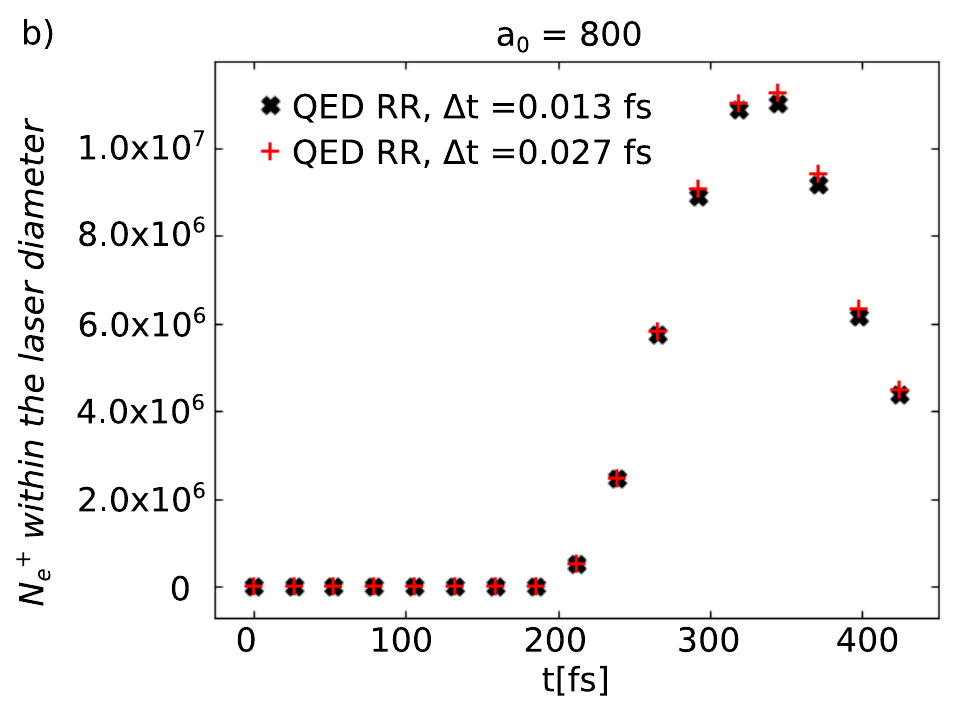}}
\caption{Time evolution of the number of positrons ${N_e}^+$sustained in the area of the laser pulse diameter $d_L$, using the parameters in section \ref{subsec:IIIA} for a) $a_0=600$ and b) $a_0=800$, for simulation time from $t = 0 $ to $t =0.43~\mathrm{ps}$.
We use a timestep of $\Delta t=0.013~\mathrm{fs}$ (black crosses) and $\Delta t=0.027~\mathrm{fs}$ (red pluses) with the Boris pusher algorithm. The simulations are carried out with the quantum radiation reaction (QED RR) regime. The interaction is placed in a vacuum.
}\label{fig:appendix1}
\end{figure}

In the appendix, we show that our simulation results converge with a decreasing timestep. For PIC simulations with high-intensity fields using the standard Boris pusher\cite{boris1970relativistic}, which we also applied in our simulations, the numerical correctness highly depends on the choice of a timestep \cite{arefiev2015temporal, vranic2016classical, gordon2017pushing, robinson2019extreme, tangtartharakul2021particle, li2021accurately}. Here, we focus on a number of positrons sustained in the laser diameter area (i.e. deflected positrons) since this is a crucial outcome of our investigation. We examined the results for the highest laser potentials we utilized, i.e. $a_0=600$ and $a_0=800$. These cases would most likely be affected by numerical errors since the higher $a_0$ the smaller $\Delta t$ is necessary \cite{arefiev2015temporal}. The results can be seen in figure \ref{fig:appendix1}, for a) $a_0=600$ and b) $a_0=800$. 
We used the same simulation parameters as described in Section \ref{subsec:IIIA} for the QED radiation reaction model, vacuum case. We decreased the original timestep  of $\Delta t=27$~as by a half and checked the time evolution of the number of positrons situated in the area of the laser diameter during the positron creation and deflection.

There are negligible differences and the results converged well with a decreased $\Delta t$. Therefore, we can assume that our general conclusions about the positron deflection were not essentially affected by the choice of a timestep and potential consequent numerical errors.

\nocite{*}
\bibliography{literature.bib}
\end{document}